# Analysis of the benefits of designing and implementing a virtual didactic model of multiple choice exam and problem-solving heuristic report, for first year engineering students.

**Abstract**. Improvements in performance and approval obtained by first year engineering students from University of Concepcion, Chile, were studied, once a virtual didactic model of multiple-choice exam, was implemented. This virtual learning resource was implemented in the Web ARCO platform and allows training, by facing test models comparable in both time and difficulty to those that they will have to solve during the course. It also provides a feedback mechanism for both: 1) The students, since they can verify the level of their knowledge. Once they have finished the simulations, they can access a complete problem-solving heuristic report of each problem; 2) The teachers, since they can obtain information about the habits of the students in their strategies of preparation; and they also can diagnose the weaknesses of the students prior to the exam.

This study indicates how this kind of preparation generates substantial improvements on the approval rates by allowing the students: 1) A more structured and oriented systematic study, and 2) To learn, from the feedback report of each problem that the information described in their exams should have, at each stage, a proper logical connection. Likewise, this study allows differentiating how the repetition in the use of the platform and the formality in the time invested indicate clear differences on the obtained performance.

1. **Introduction**

The low passing rates in the academic performance of first year physical-courses are an issue of great importance for the educational development at University of Concepcion (UdeC), Chile, among others Chilean higher education institutions. Similar problem can be found in many countries, where currently their higher education institutions not only focus on research in science and technology, but also dedicate time and resources to the improvement of science education (Chang and Chiu 2005).

The set of factors involved in the low performance and desertion students from the subjects, in many cases, include reasons external to the educational institution. Among the main causes of purely educational origin, the change in the type and criteria of exams compared to those that students usually face throughout school, could be mentioned.

The typical model of exam used at the UdeC is made of a mixture of multiple-choice including problem solving. Those types of question show different benefits. The multiple-choice instruments allow identifying student´s conceptions and have the potential to make a valuable contribution to the field of assessment (Treagust 1985, Haslam and Treagust 1987, Odom and Barrow 1995). On the other hand, problem-solving plays a crucial role in the science curriculum and instruction in most countries, and is reported by many authors as a very difficult task for students (Mettes 1980). These is a result of the lack of alternatives to practice and rehearse those modalities of exams and thus structure a systematic study, and control the anxiety in which students are involved at the time of solving their tests (Udo 2001, Mallow 2010). In order to increase the conceptual understanding of the problem, it is necessary to teach students an organized problem-solving approach that explicitly shows them all the steps involved in the problem-solving process to help them address new problems in a systematic manner (Lorenzo 2005, Maloney 1994).

In order to improve the preparation of the students, substantial improvements on the approval rates such results, and to obtain indicators on how the responsibility by which a subject is confronted affects the outcomes (Newble 1983), a virtual didactic model of multiple-choice exam with a problem-solving heuristic report was implemented in a web platform (Bradley 2002). The Web-based education environments have become important aspects of the development of the universities around the world, and it is increasingly used as a vehicle for flexible learning, where learning is seen to be free from time, geographical, and participation constraints. In addition to flexibility, the Web facilitates student-centered approaches, creating a motivating active learning environment, and provides an excellent environment for interaction between teachers and students (Own 2006, Pelton 1996, Wu 2004).

In this work the results of the analyses obtained after implementing the learning resource in the ARCO platform in real conditions is presented. This environment allowing the students: 1) A more structured and oriented systematic study, and 2) To learn, from the feedback report of each problem that the information described in their exams should have, at each stage, a proper logical connection. A lack of coherence in the development of the problems was historically the main cause for rejection, in the evaluations process. In that sense, great improvements in both percentage of approval obtained by those who systematically study by means of the platforms, and in the qualification marks, were obtained.

Moreover, all of the student´s activities are recorded, in that sense, this tool provides a feedback mechanism for the teachers, since they can obtain information about the habits of the students in their strategies of preparation; and they also can diagnose the weaknesses of the students prior to the exam. In that way, a diagnosis independent tool is available for teachers. Small changes in the habits of the students could be influenced, by assignation of tasks sequentially arranged in time.

The main contents in the platform were based on instructional concepts, allowing the students a more structured and oriented systematic study. Also, this practice environment allows the students become familiar with the specific scientific language, which is suddenly required at the entrance of the university. But we also tried to include constructivist ideas, e.g. adding more emphasis to student´s problem solving in everyday context, and taking the information of the platform in order to modify the dynamic of the classroom interactions.

We consider that education will be highly increased if knowledge and experience move forward simultaneously, in spite of quantity and complexity. In this learning resource we present only a specific set of tools that we believe are implicit in the learning process, so that the students can continue the endless process to refine, enrich and enhance their skills.

2. **Research method**

The research reported here is a case study that aimed to develop, implement a virtual didactic model of multiple-choice exam (of equal conditions to those developed during the course) which including a complete problem-solving heuristic report for each question of the exam. The study was conducted and put into action in two separated sections of students that take the same course. The universe of student population for whom the platform was available reached 190 students approximately, distributed in two sections of independent lectures, every one of 95 students approximately.

**Development of the Multiple-Choice Exam and the Problem-Solving Heuristic Report**

The virtual model of multiple-choice exam was developed and put into operation during the regular semesters of first years of engineering careers. In this virtual model of exam, the students had the chance of performing test simulations, facing during their period of study to questions and situations similar to those that they would find during the official exams (Alfred and Rovai 2004, Mattheos 2001). The platform also provides a feedback mechanism for the students, since they can verify the level of their knowledge, once they have finished the simulations – the solution of the exam is released once the simulation is finished. In this sense, this web-tool helps to the students to auto-arrange their capacities of self-learning. The platform generating at the end, the score obtained by the student and it shows a report with the correct answers and a problem-solving heuristic, entering previously by the professor.

The platform allowed store information regarding the time of admission, in order to know how far in advance the preparation for the exams was started and how much dedication was given to the study (Stewart 1997), in order to compare the results obtained in the official test.

These results were compared in both percentage of approval respect to those who did not use the platform or did not give enough seriousness to its utilization as in the performance obtained (score obtained in the exam) (Sahin 2010).

### 3. Technical Support of the Virtual Resource

The operability over which the virtual exams were mounted was in charge of the ARCO platform system, which is an online teaching support channel developed by the CFRD institution at the University of Concepción. This platform can be accessed previous authorization through the Web address: www.arco.cfrd.cl. The platform also stores information on date and time in which a student responds the test model, as well as the duration of the exercise.

#### Criteria of applicability

The virtual platform developed was enabled to be used by the students between 10 and 15 days before the official dates of every one of the tests in the course. In order to assess the performance of the students and exclude from the analysis that did not use the platform seriously, the following post-selection criteria were applied:

1) Time involved in the development of the test models must imply at least 50% of the maximum time available for the carrying out (approx. 55 minutes out of the 110 minutes available).

2) Students must perform the simulation at least 12 hours before to the exam, in order to discard those students who were practicing the morning prior or moments before.

**Parameter of evaluation of the improvement by the resource development**

The study was focused on two groups, the first who used and registries theirs results the platform and had access to a complete problem-solving heuristic report as a method of study. The second group consists of those who do not access to the platform and that formed part of the course. Finally the result of each group is compared with the final results of the total universe of student that form the independent lectures. According to selection criteria defined in the experimental proposal, the academic performance obtained by the different group. The comparison criteria were the passing rates obtained by each group and the average score obtained.

4. **Analysis and results**

The analysis allows observing great differences in the results obtained in the first exam respect to the second one. Results from the first exam indicate a high rate of passing (88% passed the exam), clearly showing that facing for the first time this kind of exams produces problems in the performance of the students. In particular, anxiety and apprehension experienced by the students in their first exam and the many times wrong way of planning the topics to study produce a low performance.

Table 1 presents a summary of the results obtained in the first exam of the course. If we analyze the low passing rate (12%) and identify to which group of students belong those who passed the exam, we can observe that of the 42 approved students from a universe of 192, 57% correspond to students that correctly used the platform (a total of 24 students) and 43% correspond to students who did not use the platform (18 students).

On the other hand, if we analyse the passing percentage obtained by every studied group, we found that who used the platform (a universe of 60 students), 40% of them could pass the first exam (24 students), whereas from the group that did not use the platform (a universe of 131 students), only 12% could pass such exam (Fig. 1).

If we analyze the average score obtained by each group, we can observe that the average of the students under study was 3.1 (with 4.0 the minimum passing mark). From this result, it can be observed that the averages obtained by each group are: who used correctly the platform had an average score of 3.6 (which is lower than the general average of the course). On the other hand, the group formed by those who did not use the platform had and average of 2.6, which id 16% lower than the general average of the course and 28% lower than the average of those who used the virtual platform (Fig. 2).

|  | **Group that did not use the platform** | **Group that used the platform** | **Results of the studied course** |
|---|---|---|---|
| **Total number of students in the study** | **131** | **60** | **191** |
| **Number of reprobate students** | **115** | **36** | **149** |
| **Number of students approved** | **18** | **24** | **42** |
| **Reprobate percentage** | **88%** | **60%** | **78%** |
| **Approved percentage** | **12%** | **40%** | **22%** |
| **Average** | **2,6** | **3,6** | **3,1** |

**Table 1. Summary of the results obtained in the first exam of the course**

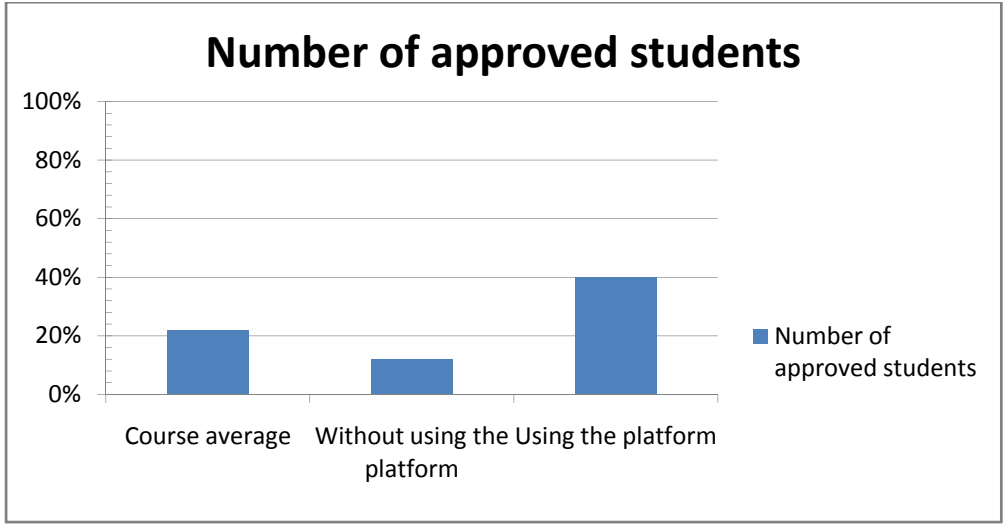

**figure 1. Comparative among the passing percentage obtained by every studied group (the course, the students which without using the platform and the student which using the platform)**

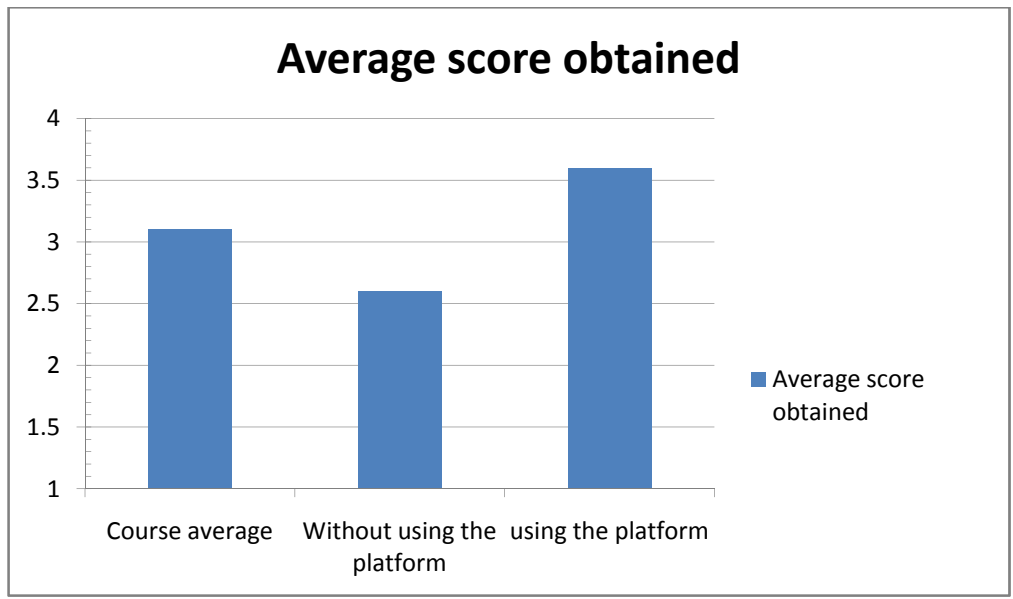

**Figure 2. Comparative among the average score obtained by every studied group in the first exam (the course, the students which without using the platform and the student which using the platform)**

Finally, a comparative between the results obtained in the platform respect to the score obtained in the exam are shown in the Fig 3. The distribution in the qualification obtained for the students, which used the platform; show a similarity for the first exam.

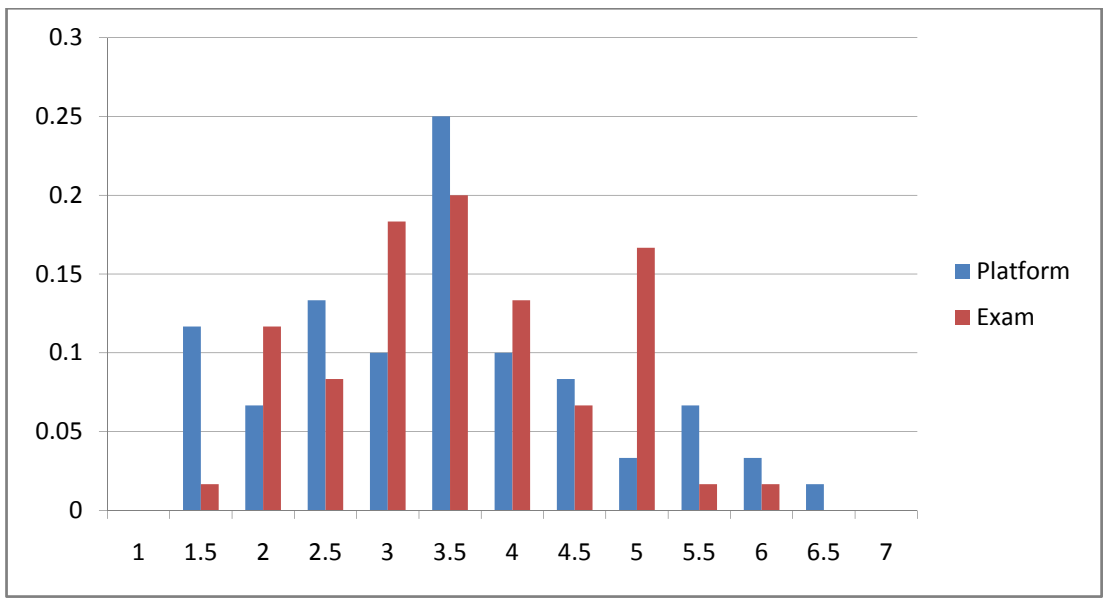

**Figure 3. Comparative between the results obtained in the platform respect to the score obtained in the first exam for the students which used the platform.**

For the second exam a radical change was observed in the performance of the courses results. This exam shows a great rise in the passing rate of the course (65% passed the exam). This rate corresponds to an increment of almost 4 times the amount of passing students respect to the first exam.

This is a clear indicator that after being faced for the first time to this kind of exam. Students are able to better face the second exam, generating better strategies of study and distributing in a better way the time to develop each problem.

Table 2 presents a summary of the results obtained by the students in the second exam of their course. If we analyze the passing percentage obtained by each group we can see that who used the platform (a universe of 38 students), 68% could pass their first exam (19 students), whereas in the group that did not use the platform (a universe of 162 students), 64% of them could pass such test (Fig. 4).

|  | Group that did not use the platform | Group that used the platform | Results of the studied course |
|---|---|---|---|
| **Total number of students in the study** | 162 | 28 | 190 |
| **Number of reprobate students** | 58 | 9 | 67 |
| **Number of students approved** | 104 | 19 | 123 |
| **Reprobate** |  |  |  |
| percentage | 36% | 32% | 35% |
| Approved percentage | 64% | 68% | 65% |
| Average | 4,1 | 4,6 | 4,2 |

Table 2. Summary of the results obtained in the second exam of the course

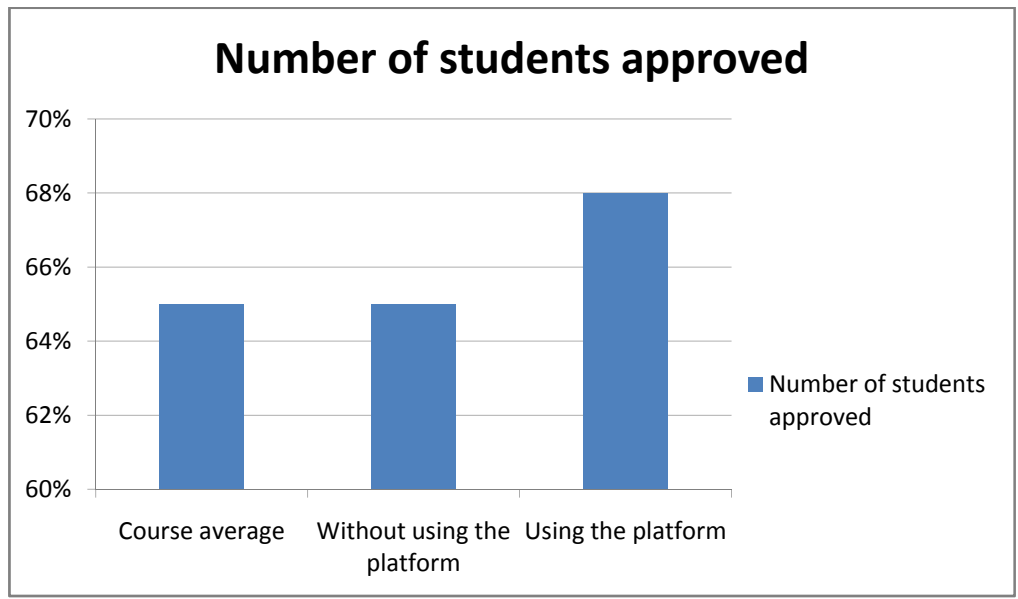

Figure 4. Comparative among the passing percentage obtained by every studied group in the second exam (the course, the students which without using the platform and the student which using the platform)

From this result, if we study the average score obtained by each group, we can observe that the average from the universe of studied students was 2.1 (being 4.0 the minimum passing mark). From this result it can be observed that the averages obtained by each group are:

Who correctly used the platform obtained an average score of 4.6 (which is 9% higher than the general average of the course). On the other hand, the group formed by who did not use the platform had an average of 4.1, which is 2.3% lower than the general average of the course and 11.8% lower than the average of those who used the virtual platform (Fig. 5).

Finally, a new comparative between the results obtained in the platform respect to the score obtained y en exam are shown in the Fig 6. The distribution in the qualification obtained for the students, which used the platform; show a high increment in the score obtained in the exam after used the platform.

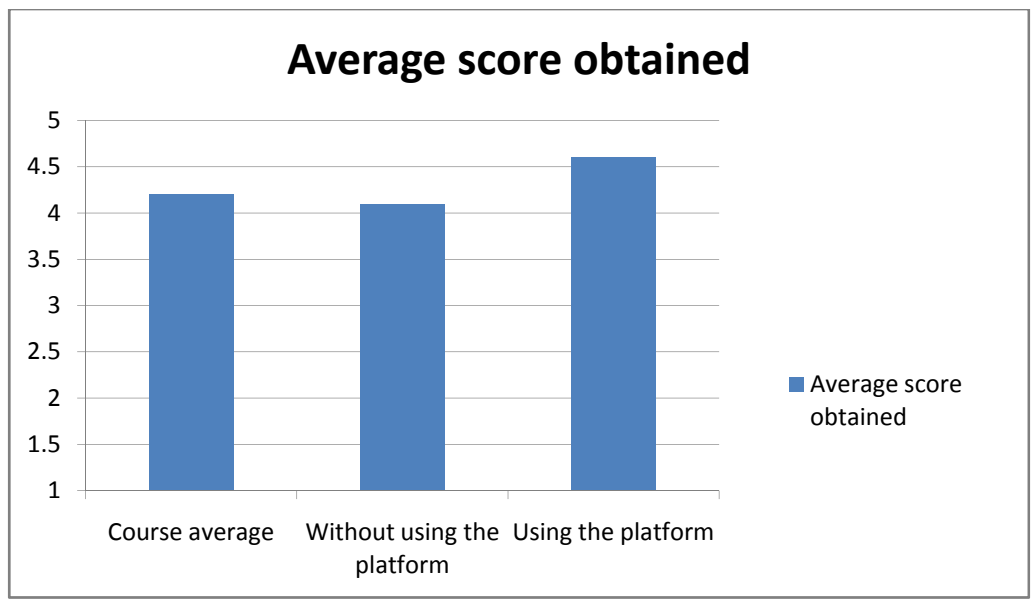

**Figure 5. Comparative among the average score obtained by every studied group in the second exam (the course, the students which without using the platform and the student which using the platform)**

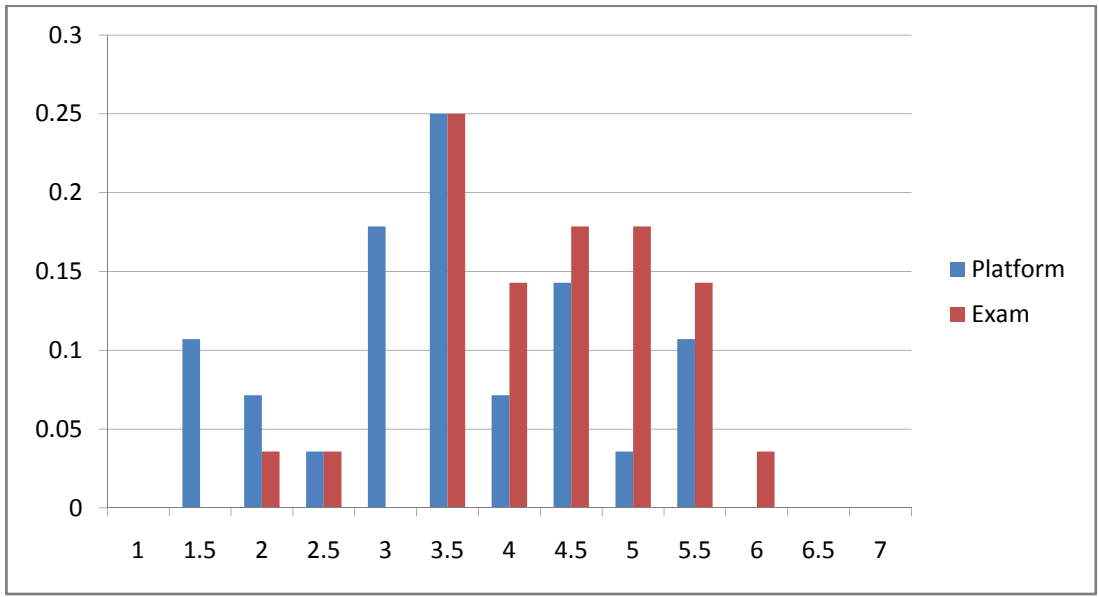

**Figure 6. Comparative between the results obtained in the platform respect to the score obtained in the second exam for the students which used the platform.**

5. Conclusions

The present study involved the development and implementation of a virtual simulation platform, by which the students could prepare the study of their exams, systematically and with days of advance. They could also practice, facing actual mock exams, with problems and questions of the equivalent level and at the same time, available in real conditions. Moreover, all of the student´s activities are recorded, in that sense, information about the habits of the students can be obtained. The study allowed observing along a regular academia period, how the introduction of new tools allow a trained in the development and a complete solving of exams, showing clearly increasing in the performance levels and passing rates.

This virtual learning resource provides a feedback mechanism for both: 1) The students, since they can verify the level of their knowledge. Once they have finished the simulations, they can access a complete problem-solving heuristic report of each problem; 2) The teachers, since they can obtain information about the habits of the students in their strategies of preparation; and they also can diagnose the weaknesses of the students prior to the exam. Also, small changes in the habits of the students could be influenced, by assignation of tasks properly arranged in time.

The increase in academic performance of those who correctly used the platform and the normal evolution of this academic performance among those who did not use the platform throughout the course reflected as a key component for the low scores is the inexperience of coping exams of this modality. This affects with anxiety and incapacity of facing properly the evaluating instrument. Availability of this platform and the fact of offer the students to test in quiet as well as in advance, generate a better study in students, clearly indicating them which are the subjects in which they are weaker and allowing them coping in a better way the official exam. Also, this environment allowing the students to learn, from the feedback report of each problem that the information described in their exams should have, at each stage, a proper logical connection. A lack of coherence in the development of the problems was historically the main cause for rejection, in the evaluations process. In that sense, we have observed a strong correlation between a proper use of the platform and the obtained results.

The main contents in the platform were based on instructional concepts, allowing the students a more structured and oriented systematic study. Also, this practice environment allows the students become familiar with the specific scientific language, which is suddenly required at the entrance of the university. But we also tried to include constructivist ideas, e.g. adding more emphasis to student´s problem solving in everyday context, and taking the information obtained from the platform in order to modify the classroom interactions.

In the near future, we look for the implementing simple virtual laboratory tasks, where students can be trained to obtain and process information. In these virtual experiments, they would be required to work in groups, in order to increase their interactions.

**Acknowledgements**. This work was supported by the Dirección de Docencia, at Universidad de Concepción, through project no. 12-012.

(4), 297-314

18. Sahin, M. (2010) Effects of Problem-Based Learning on University Students' Epistemological Beliefs About Physics and Physics Learning and Conceptual Understanding of Newtonian Mechanics. Journal of Science Education and Technology. 19 (3) 266–275.